\documentclass[11pt]{article}
\usepackage{geometry}                
\usepackage{graphicx}
\usepackage{amssymb}
\usepackage{epstopdf}
\title{ Magnetic Yang-Mills Theory as an Effective Field Theory
of the Gluon Plasma}
        
\author{M. Baker\\  {\it Department of Physics, University of Washington} \\  {\it  Box 351560, Seattle WA 98195, USA}}
\date{}
\begin{document}
\maketitle
\begin{abstract}
                              
We propose magnetic $SU(N)$ pure gauge theory as an
effective field theory describing the long distance nonperturbative 
magnetic response of the deconfined phase of Yang-Mills theory.
The magnetic non-Abelian Lagrangian, unlike that of electrodynamics 
where there is exact electromagnetic duality, is not known explicitly,
but here we regard the magnetic $SU(N)$ Yang-Mills Lagrangian as the 
leading term in the long distance effective gauge theory of the plasma
phase. In this treatment, which is applicable for 
a range of temperatures in the interval $T_c\, < \,T \, < \,3 \,T_c$ 
accessible in heavy ion experiments, formation of the magnetic energy 
profile around a spatial Wilson loop in the deconfined phase parallels 
the formation of an electric flux tube in the confined phase. We use 
the effective theory to calculate spatial Wilson loops and the  
magnetic charge density induced in the plasma by the corresponding 
color electric current loops. These calculations suggest that the deconfined phase of Yang-Mills theory has the properties of a closely-packed fluid of magnetically charged composite objects.
\end{abstract}


\section{Introduction}

The confined phase of $SU(N)$ Yang-Mills theory can be described by an 
effective theory coupling
magnetic $SU(N)$ gauge potentials ${\bf C}_\mu$ to three 
adjoint representation Higgs fields  \cite{BBZ1991}.  
In the confined phase the magnetic gauge symmetry is completely 
broken via a dual Higgs mechanism in which all particles become 
massive.  The value $\phi_0 $ of the magnetic  Higgs condensate 
is determined by the location of the minimum in the Higgs potential, 
and the dual gluon acquires a mass 
\begin{equation}
M_g \sim g_m \phi_0  \, ,
\label{dualgluon}
\end{equation}
where $g_m$ is the magnetic gauge coupling constant.
The coupling of the potentials ${\bf C}_\mu$ to the magnetically
charged Higgs fields generates color magnetic currents which,
via a dual Meissner effect,
confine $Z_N$ electric flux to narrow tubes connecting a
quark-antiquark pair  \cite {Mandelstam}.
For $SU(3)$, the dual gluon mass
 $M_g  \,\sim \, 1.95 \sqrt {\sigma} $ \cite{BBZ1997}.
The effective theory is applicable at distances greater 
than the flux tube radius  $R_{FT} \sim \frac{1}{M_g} \sim 0.3 fm $.  
Since $SU(3)$ lattice simulations \cite{wenger2002} 
yield a  deconfinement temperature 
$T_c \,\approx \, 0.65 \sqrt { \sigma}$, the scale
$M_g \,\sim \, 3T_c$.  
Thus there is a range of temperatures  within the interval $T_c\, < \, T\,< 3\,T_c$ where the effective theory 
should also be applicable in the deconfined phase.

\section{Effective Magnetic Yang-Mills Theory of the Deconfined Phase.}
\label{sec:deconfined}

Above $T_c$ the Higgs condensate vanishes, so the magnetic gluon 
becomes massless.  We assume that the Higgs fields do not play an 
essential role in the deconfined phase \cite{baker}. The effective
theory then reduces to  pure SU(N) Yang-Mills theory of magnetic gauge 
potentials $ { \bf C}_\mu \equiv ({\bf C}_0, { \vec  {\bf C} })$. 
This theory has the same form as the microscopic 
electric theory, but with a fixed gauge coupling 
constant $g_m \,\sim \, 3.91$ and  fixed ultraviolet cutoff 
$M_g \, \sim \, 3T_c \, \sim \, 800 MeV $, values 
determined by the effective theory description
of the confined phase.  The resulting long wavelength magnetic gluons, 
which at $T=0$ confine $Z_N$ electric flux,  
are the elementary degrees of freedom for $T\, >\, T_c$.

We regard magnetic $SU(N)$ gauge theory
as an effective field theory appropriate for calculating
the long distance magnetic response of the gluon plasma.
The dual effective Lagrangian $L_{eff} ({\bf C}_\mu)$ contains all 
combinations of $ {\bf C}_\mu $ 
invariant under magnetic non-Abelian gauge transformations:
\begin{equation}
L_{eff} ({\bf C}_\mu) = 2 tr \left[-\frac{1}{4} {\bf G^{\mu \nu} G_{\mu \nu}} + \cdots \right] \, ,
\label{Leff}
\end{equation}

where

\begin{equation}
{\bf G_{\mu \nu}} = \partial_\mu {\bf C_\nu} - \partial_\nu {\bf C_\mu} - i g_m [ {\bf C_\mu, C_\nu}] \, . 
\label{gmunu}
\end{equation}
Here we retain only the leading term in $L_{eff} $,
pure magnetic Yang-Mills theory.

We use effective magnetic Yang-Mills theory to 
calculate spatial Wilson loops measuring $Z_N$ magnetic flux
$k$ passing through a loop $L$.
These loops are  obtained from the partition function of the 
magnetic theory in the presence of a current of $k$ quarks 
circulating around the loop $L$, producing  a steady current
$\frac{ 2 \pi T}{g_m} {\bf Y}_k$, 
where ${\bf Y}_k$ is a diagonal matrix with the property
$e^{2\pi i {\bf Y}_k} = e^{2\pi i \frac{k}{N}}$ \cite{giov}. 
This current is the source of an external 
magnetostatic scalar potential, ${\bf C}_0^{ext} =
\frac{2 \pi T}{g_m} {\bf Y}_k \frac {\Omega_S ({\vec x})}{4 \pi}$, 
where $\Omega_S ({\vec x})$ is the solid angle subtended 
at the point ${\vec x}$ by a surface $S$ bounded by the loop $L$.
Then $\vec \nabla {\bf C}_0^{ext} = \vec {\bf B}_{BS}$, the 
Biot-Savart magnetic field of the current loop.
(The color magnetic field $ { \vec {\bf B}_j} = {\bf G}_{0j}$.) 
The spatial Wilson loop of Yang-Mills theory, calculated in the
effective magnetic gauge theory, is then the partition function
of the magnetic theory in the presence of the 
external potential ${\bf C}_0^{ext}$.

\section{The Spatial Wilson Loop Calculated in the Magnetic Theory.}
\label{sec:wilson}

To evaluate the partition function of the magnetic theory
in the deconfined phase, where there is no Higgs potential,
we must calculate the one loop effective potential $U({\bf C}_0)$ 
of magnetic Yang-Mills theory
in the background of a static magnetic scalar potential ${\bf C}_0$:

\begin{equation}
e^{-\int {d{ \vec x} \frac {U({\bf C}_0)}{T}}} \equiv 
e^{-S^{1-loop}({\bf C}_0)} 
=  Det \left (- D^2_{adj} ({\bf C}_0) \right) \, .
\label{Z1loop}
\end{equation}
$U({\bf C}_0)$ is the
counterpart in the deconfined phase of the classical Higgs
potential generating electric flux tube solutions in the confined
phase, and gives rise to the spontaneous  
breakdown of the $Z_N$ symmetry of the effective magnetic gauge theory.
We evaluate $U({\bf C}_0)$ integrating over the massless modes of 
magnetic Yang-Mills theory, introducing a Pauli-Villars regulator 
mass $M$. This mass should approximately be equal to the 
dual gluon mass $M_g$ determining the maximum energy of the modes
included in the effective theory.
Aside from the presence of the regulator, the calculation of 
$U({\bf C}_0)$ mimics the calculation  \cite{gross,gocksch}  of the one loop effective 
potential $U({\bf A}_0)$ in Yang-Mills theory used \cite{kovner,giov3} to calculate 
 the spatial 't Hooft loop \cite{thooft,philippe}
. 
We assume that the background potential 
${\bf C}_0$ has the same Abelian color structure 
as ${\bf C}_0^{ext}$, i.e.,
${\bf C}_0  =  \frac {2 \pi T}{g_m} C_0  (\vec x)  {\bf Y}_k$. 
The corresponding dimensionless effective potential 
$U(C_0, \frac{T}{M})$ 
is then a periodic function of $C_0$ with period $1$.
The resulting expression for the one loop effective 
action $S^{1-loop} (C_0)$ is given by \cite{baker}
\begin{equation}
S^{1-loop}( C_0)  = \frac { 4 \pi^2 \sqrt 3 \, k (N-k) }{N^{3/2}g_m^3} \int {d{\vec x} U( C_0,\frac{T}{M})} \, ,
\label{S1loop}
\end{equation}

where

\begin{equation}
 U (C_0,\frac{T}{M}) =  \left [ {[C_0 ]}^2(1 - [C_0 ])^2  - \frac{3}{4 \pi^4} I ( C_0, \frac{T}{M}) \right]  \, ,
\label{U}
\end{equation}

and

\begin {equation}
I ( C_0, \frac{T}{M})  = \int_0^\infty { dy \,\,y^2 log \left (\frac{cosh \sqrt{y^2 + (\frac{M}{T})^2} - cos (2 \pi C_0)}{cosh \sqrt{y^2 + (\frac{M}{T})^2} -1 } \right)}  \, ,
\label{IC0}
\end{equation}
with $[C_0 ] \equiv {|C_0 |}_{mod1}$. 
The integral $I (C_0, \frac{T}{M})$ 
suppresses the short distance contribution to
 $U(C_0)$.

We separate the background scalar potential $C_0$ into the contribution 
$\frac {\Omega_S}{4 \pi}$ of the external potential
and a remaining contribution $c_0$ whose sources are the magnetic
charges of the plasma:
\begin{equation}
C_0 = c_0 + \frac {\Omega_S}{4 \pi} \,.
\label{c0}
\end{equation}
Then making the replacement (\ref{c0})
in $S^{1-loop}$ and adding the magnetic energy $( \vec \nabla c_0)^2$ 
of the induced magnetic charges
gives the effective action $S_{eff} (c_0)$: 
\begin{equation}
S_{eff}( c_0)  = \frac{ 4 \pi^2 \sqrt 3  \,k (N-k)}{N^{3/2}g_m^3} \int {d {\vec x} \left [ ( \vec \nabla  c_0  ) ^2 + {U}( c_0 + \frac  {\Omega_S}{4 \pi} )  \right ]} \, .
\label{Seff}
\end{equation}
The action (\ref{Seff}) generates a mass scale $\frac{Ng_mT}{3}$ 
and a corresponding distance scale which is used in Eqs. (\ref{S1loop}) 
and (\ref{Seff}).  For $T \, > \, T_c$ the scale $\frac{Ng_mT}{3}$ 
is greater than the cutoff $M$ so that
we can use $S_{eff}$ at the classical level to determine the leading long distance behavior of spatial Wilson loops in the deconfined phase in the same manner that the classical gauge-Higgs action was used to evaluate temporal Wilson loops in the confined phase.

The minimization of $S_{eff} (c_0)$  yields "Poisson's equation" for $c_0$:
\begin{equation}
 \nabla^2 c_0 (\vec x) = \rho_{mag} (\vec x) \, ,
\label{poisson}
\end{equation}
where
\begin{equation}
\rho_{mag} (\vec x) = \frac{1}{2} \frac{dU(c_0 + \frac{\Omega_S}{4 \pi} ,\frac{T}{M})}{d c_0}    
\label{magneticcharge}
\end{equation}
is the color magnetic charge density induced in the vacuum by the
current loop. 
The boundary conditions on $c_0$ are:
for $\vec x$ on $L$, $c_0 (\vec x) \rightarrow 0$, and 
as $\vec x \rightarrow \infty$,
$c_0 (\vec{x}) \rightarrow - \frac {\Omega_S (\vec x)}{4 \pi }$.
The latter condition means that the induced magnetic charges
screen the external field
so that the total field
 $\vec B(\vec x) = \vec \nabla {c_0} + \vec B_{BS} $ 
has an exponential falloff determined by the "Debye" magnetic
screening mass $m_{mag} (T)$. In unscaled units
\begin{equation}
m_{mag} ^2(T) =  \frac{N {g_m}^2 T^2 }{6} \frac{ d^2 U( C_0,\frac{T}{M})}{ d C_0^2} \bigg|_{C_0 = 0 } \, .
\label{magneticmass}
\end{equation}
The value $S_{eff} (L)$  of $S_{eff}(c_0)$ at its minimum
determines the spatial Wilson loop $W(L) = e^{-S_{eff} (L)} $.
  \begin{figure}[ht] 
   \centering
     \includegraphics[width=5in]{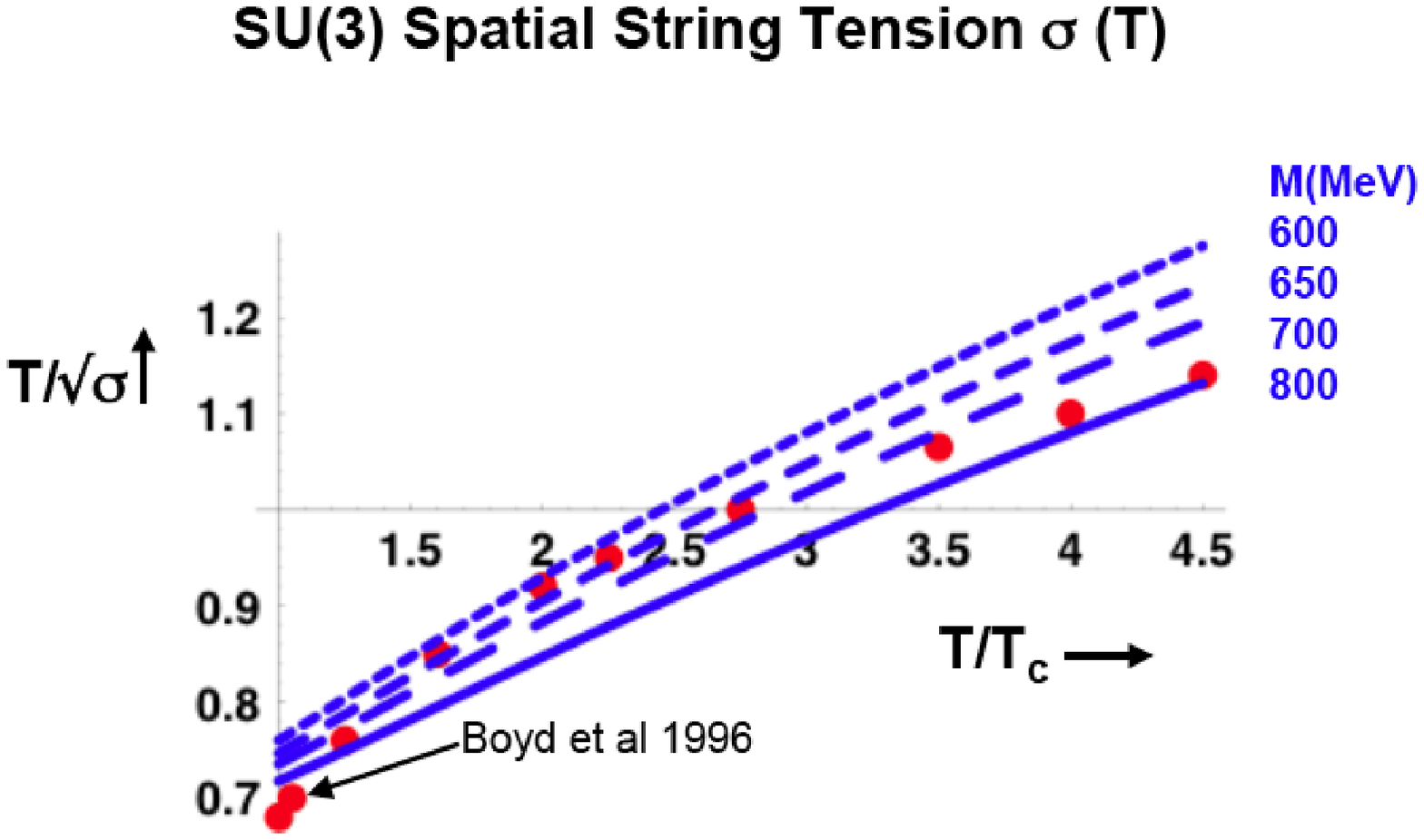}
       \caption{  Comparison of $SU(3)$ 4d lattice data (dots) \cite{boyd1996} for the spatial string tension $\sigma(T)$ with the prediction  of the effective  magnetic Yang-Mills theory for four values of the Pauli-Villars regulator mass $M$.}
 \label{fig:fig7}
\end{figure}
  \begin{figure}[ht] 
   \centering
     \includegraphics[width=5in]{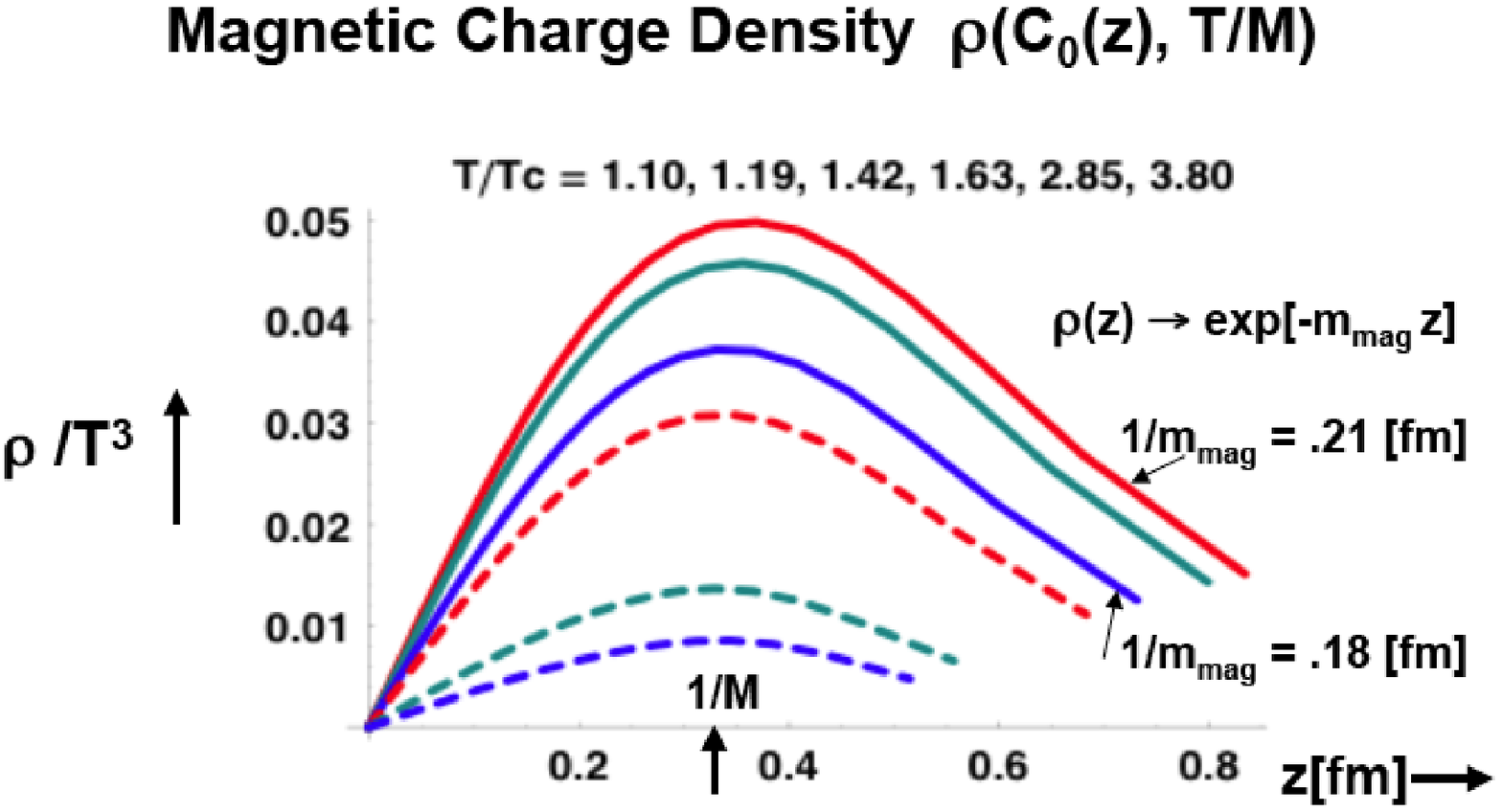}
       \caption{ Induced magnetic charge densities $\rho (z)$ outside a 
large Wilson loop for varying temperatures.} 
         \label{fig:fig9}
\end{figure}

\section{Spatial String Tension and Induced Magnetic Charge Density}

As $  L \rightarrow \infty$  ,
$S_{eff} (L) \rightarrow L^2 \sigma_k (T)$, 
determining the spatial string tension $\sigma_k (T)$.
In this limit the solid angle $\Omega_S = -2 \pi$  for $z>0$ 
and $ 2 \pi$ for $z<0$, where $z=0$ is the plane
of the loop $L$. The boundary condition at large 
distances becomes $c_0 \rightarrow \pm \frac{1}{2}$ as 
$z \rightarrow \pm \infty$, and
$c_0$ becomes a function only of $z$.
Evaluating $S_{eff} (c_0)$ at the  "classical" solution $c_0 (z)$ 
yields: 
\begin{equation}
\frac{\sigma_k (T)}{T^2}=\frac{8\pi^2 k(N-k)
 \int_{0}^{1} dC_0 
\sqrt {U(C_0,\frac{T}{M})} 
}{ g_m{\sqrt{3N}}} \, .
\label{sigmakt}
\end{equation}

Eq. (\ref{sigmakt}) is valid for any $SU(N)$ group, but the values of $g_m$ and $M_g$ have been determined only for $SU(3)$ where the effective theory has been applied in the confined phase.  
The temperature dependence of  the ratio $\frac{\sigma_k (T)}{T^2}$
comes from the Pauli-Villars cutoff, which suppresses the contributions 
of momenta greater than $M$ to the effective potential (\ref{U}) and
consequently to $ \sigma_k (T) $. 
Since the Pauli-Villars regulator is rather "soft", 
allowing substantial contributions from momenta greater than $M$, 
we have also evaluated the string 
tension using values of $M$ smaller than $M_g \sim 800 MeV$.
In Fig. \ref{fig:fig7} 
we plot $\frac{T}{\sqrt{\sigma (T)}}$, Eq. (\ref{sigmakt}) for $SU(3)$, 
as a function of $\frac{T}{T_c}$ for a range of values of
$M$ and compare with the results of 4d lattice 
simulations \cite{boyd1996}.  
(Values of $M$ lying between $600 MeV$ and $650 MeV$ 
give the best fits to the lattice data in the temperature 
interval $T_c\, <\, T\, < \, 3 \, T_c $.)

In Fig. \ref{fig:fig9} we plot, for a range of temperatures using
$M =600 MeV$, the magnetic charge density 
(\ref{magneticcharge}) induced by a large Wilson loop as a function
of the distance from the loop. For these
temperatures the induced distributions $\rho (z)$ of magnetic charge 
have maxima for values of $z$ close to $\frac{1}{M} \sim 0.33fm$. 
This can be understood since $\frac{1}{M}$ is the shortest wavelength 
of the quanta contributing to $U$  and hence determines the spatial 
extension of the magnetically charged objects described by effective 
theory. 
Their "size" $\frac{1}{M}$ is greater than the 
magnetic screening length $\frac{1}{m_{mag} (T)}$ determining the 
exponential tail of the charge distributions 
as $z \rightarrow \infty$. 

In a dilute plasma of charged ions the size of the ion cloud is 
determined by the Debye screening length which is much larger
than the mean separation between the ions. By contrast, in the 
gluon plasma 
the mean distance between the extended charges, determined by their
intrinsic size $\sim \frac{1}{M}$, is greater than the screening length
characterizing the tail of the distribution. 
This gives a physical picture of the gluon plasma as a dense (closely-packed) liquid of extended magnetically charged objects.  Comparison of the plots in 
Fig. \ref{fig:fig9} with correlation functions calculated 
in lattice simulations of Yang-Mills theory \cite{maxim,elia}
could check this picture.

\section {Summary}
\label{sec:summary}

According to our picture, the plasma 
phase of $SU(N)$ Yang-Mills theory in a temperature range included
the interval $T_c\,<\,T\,<\,3\,T_c$ 
is described by effective magnetic $SU(N)$ gauge theory. 
Integrating out the long wavelength non-Abelian degrees of freedom
gives rise to extended magnetic charges confining magnetic flux,
which are the counterpart in the deconfined phase of magnetic currents 
confining electric flux in the confined phase.
Extension of our work to calculate non-equilibrium 
quantities would make it possible to use effective magnetic 
gauge theory to analyze experiments on heavy 
ion collisions.

\section*{Acknowledgments}
I would like to thank   B. Bringoltz, M. Chernodub, Ph. de Forcrand, M. Fromm, 
 C. P. Korthals Altes, A. Vuorinen and L. Yaffe for their valuable help.

\end{document}